\documentclass[aps,prb,twocolumn,superscriptaddress,showkeys,showpacs,floatfix]{revtex4-2}
\usepackage{amsmath,amssymb,bm}
\usepackage{graphicx}
\usepackage{dcolumn}
\usepackage{booktabs}
\usepackage{multirow}
\usepackage{color}
\usepackage[dvipsnames]{xcolor}
\usepackage{hyperref}
\usepackage{txfonts}
\usepackage{comment}
\usepackage[normalem]{ulem}

\begin{document}

\title{Interplay of Magnetic Order, Structural Stability, and Orbital Ordering in BaFe$_2$$X_3$ ($X$ = S and Se)}

\author{Kunihiko Yamauchi}
\affiliation{Center for Spintronics Research Network (CSRN), The University of Osaka, Toyonaka, Osaka 560-8531, Japan}

\author{Takuya Aoyama}
\affiliation{Program of Quantum Matter, Graduate School of Advanced Science and Engineering, Hiroshima University, Higashi-Hiroshima, Hiroshima 739-8530, Japan}
\affiliation{International Institute for Sustainability with Knotted Chiral Meta Matter (WPI-SKCM2), Hiroshima University, Higashi-Hiroshima, Hiroshima 739-8531, Japan}

\author{Kenya Ohgushi}
\affiliation{Department of Physics, Graduate School of Science, Tohoku University, Sendai 980-8578, Japan}

\date{\today}

\begin{abstract}

BaFe$_2$$X_3$ ($X$ = S and Se) are quasi-one-dimensional Mott insulators with a ladder structure that exhibit Stripe- and Block-type antiferromagnetic order, respectively. The ladder arrangement of Fe atoms and strong electron correlations give rise to rich magnetic and orbital phenomena, including pressure-induced superconductivity and orbital-selective electronic states. Both compounds also show resistivity anomalies above the Néel temperature, suggesting that orbital degrees of freedom play an important role in their electronic properties. To clarify the interplay among magnetic order, structural stability, orbital ordering, and transport properties, we performed resistivity measurements and first-principles calculations for BaFe$_2$S$_3$ and BaFe$_2$Se$_3$, systematically comparing candidate magnetic and crystal structures. We find that the magnetic configuration strongly influences the stable crystal structure, orbital ordering, and transport anisotropy, providing a microscopic understanding of the contrasting electronic properties of the two compounds.

\end{abstract}

\maketitle

\section{Introduction}
The ladder-type iron-based compounds $A$Fe$_2X_3$ ($X$ = S and Se) constitute a family of quasi-one-dimensional systems that exhibit pressure-induced superconductivity.~\cite{Takahashi2015, Ying2017}
In these materials, Fe atoms are tetrahedrally coordinated by chalcogen anions, and the edge-sharing Fe$X_4$ tetrahedra form a two-leg ladder structure.
Although the local coordination is common to two-dimensional iron-based superconductors with square lattices, their electronic properties are markedly different: whereas most two-dimensional iron-based compounds are metallic, the ladder compounds are insulating at ambient pressure, reflecting their reduced dimensionality and enhanced electronic correlations.
Furthermore, the correlation strength depends sensitively on the orbital character, and these systems have been discussed in the context of orbital-selective Mott physics.
In this paper, we refer to the characteristic directions of the ladder as the leg, rung, and layer directions.

\begin{figure}
\centering
\includegraphics[width=\columnwidth,pagebox=cropbox,clip]
{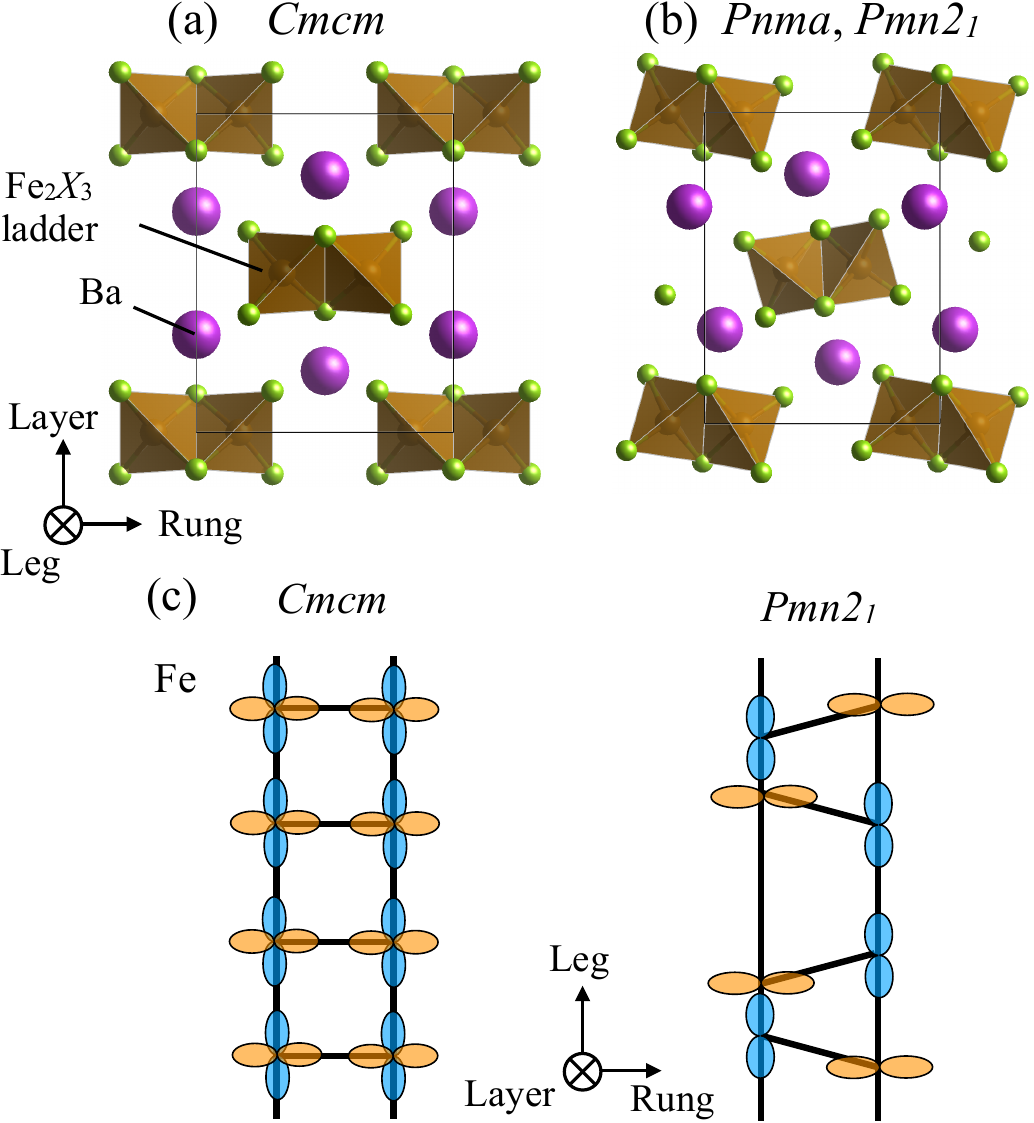}
\caption{
Crystal structure of BaFe$_2X_3$ in (a) $Cmcm$ phase and (b) $Pnma$ phase.
(c) Schematic drawings of the local ladder structure and the possible patterns of the Fe-$d$ orbital ordering with the space group of $Cmcm$ and $Pmn2_1$. 
}
\label{fig:crys}
\end{figure}

\begin{figure}
\centering
\includegraphics[width=6.5cm,pagebox=cropbox,clip]{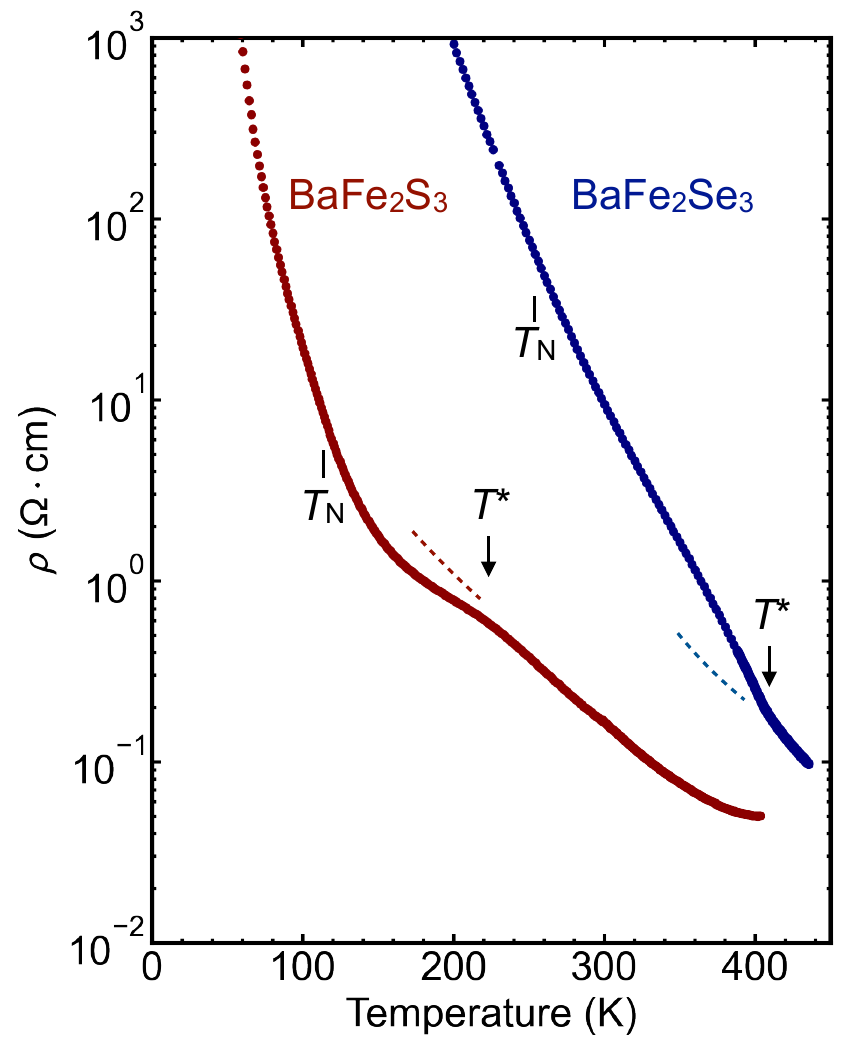}
\caption{
Temperature dependence of electric resistivity. The $T_{\rm N}$ and $T^{\star}$ indicate antiferromagnetic transition and electronic anomaly related to the orbital state.}
\label{fig:exp}
\end{figure}

BaFe$_2$S$_3$ crystallizes in the $Cmcm$ space group with a uniform ladder structure (Fig.~\ref{fig:crys}(a)).
At approximately $T^\ast \sim 200$~K, well above the magnetic transition temperature, the electrical resistivity exhibits a non-monotonic temperature dependence: below $T^\ast$, the increase in resistivity is suppressed relative to the high-temperature extrapolation, as indicated by the dotted line in Fig.~\ref{fig:exp}.
Although no accompanying crystallographic symmetry change has been established, this anomaly has been associated with an orbital-dependent electronic reconstruction~\cite{Yamauchi2015,Ootsuki2015,Takubo2017,Hosoi2020}.
Upon further cooling below $T_{\rm N}=120$ K, BaFe$_2$S$_3$ develops Stripe-type antiferromagnetic order with a magnetic propagation vector $\mathbf{q}=(1/2,1/2,0)$. 
Recent angle-resolved photoemission spectroscopy has reported momentum-dependent spin splitting in the antiferromagnetic phase, suggesting the emergence of an altermagnetic electronic structure.\cite{Iwasaki2025}

In contrast, BaFe$_2$Se$_3$ undergoes successive structural phase transitions upon cooling.
Above $T_{s1}\sim660$ K, it crystallizes in the high-symmetry centrosymmetric $Cmcm$ structure.
At $T_{s1}$, it transforms into the centrosymmetric $Pnma$ structure, accompanied by tilting and distortion of the FeSe$_4$ ladders~\cite{Svitlyk2013}.
Upon further cooling, a second structural phase transition occurs around $T_{s2}\sim400$ K, as evidenced by synchrotron x-ray diffraction and differential scanning calorimetry~\cite{Svitlyk2013}.
Optical second-harmonic-generation and neutron-diffraction measurements further revealed that the low-temperature phase is polar with the orthorhombic space group $Pmn2_1$~\cite{Aoyama2019} (Fig.~\ref{fig:crys}(b)).
More recently, the crystal symmetry has been suggested to be further reduced to the monoclinic space group $Pm$~\cite{Weseloh2022}.
In BaFe$_2$Se$_3$, the electrical resistivity is enhanced below $T^\ast$ relative to the high-temperature extrapolation~(Fig.~\ref{fig:exp}), suggesting an orbital reconstruction with an antiferroic arrangement (Fig.~\ref{fig:crys}(c)) ~\cite{Aoyama2019,Imaizumi2020}.
Upon further cooling below $T_{\rm N}\sim250$ K, BaFe$_2$Se$_3$ develops block-type antiferromagnetic order with a magnetic propagation vector $\mathbf{q}=(1/2,1/2,1/2)$~\cite{Caron2011,Nambu2012}.

Such electronic states are also supported by spectroscopic studies.
Photoemission and x-ray absorption measurements have revealed that localized and itinerant Fe $3d$ electrons coexist in both BaFe$_2$S$_3$ and BaFe$_2$Se$_3$, indicating that these compounds cannot be described as uniform Mott insulators.
The partial delocalization of the Fe $3d$ electrons has been attributed to the strong hybridization and charge transfer between the Fe $3d$ and chalcogen $p$ orbitals~\cite{Ootsuki2015,Takubo2017}.
Furthermore, polarization-dependent x-ray absorption measurements revealed pronounced electronic anisotropy in both compounds already at room temperature, suggesting the presence of orbital order or orbital fluctuations well above $T_{\mathrm N}$~\cite{Takubo2017}.
Resonant elastic x-ray scattering measurements on BaFe$_2$S$_3$ further demonstrated that the orbital-related modulation persists above $T_{\mathrm N}$, whereas the magnetic scattering is predominantly associated with the itinerant states near the Fermi level~\cite{Takubo2025}.
These results support an orbital-selective picture in which localized and itinerant electronic components play distinct roles in the magnetic and transport properties of the iron-ladder compounds.

The remaining unresolved issue is how magnetic order determines the orbital state and the resulting transport properties. 
Experimentally, the uniform ladder geometry of the high-symmetry $Cmcm$ structure is associated with Stripe-type antiferromagnetic order, whereas block-type antiferromagnetic order is observed in the distorted ladder structure.
This Stripe order consists of antiferromagnetic spin alignment along the ladder legs and ferromagnetic alignment along the rungs~\cite{Patel2016,Wang2017}, and is regarded as the quasi-one-dimensional counterpart of the Stripe antiferromagnetic order in 122-type iron-based superconductors such as BaFe$_2$As$_2$~\cite{Dai2015}.
In contrast, the distorted ladder geometry with alternating Fe--Fe bond lengths is considered to favor the formation of block-type antiferromagnetic order composed of ferromagnetically aligned Fe$_4$ spin blocks~\cite{Nambu2012,Dong2014}.
These contrasting magnetic ground states suggest that the structural stability is intimately linked to the magnetic configuration through orbital-dependent electronic states.
Clarifying this mutual interplay is therefore essential for understanding the electronic properties of iron ladder compounds.

In this study, we perform first-principles calculations for BaFe$_2$S$_3$ and BaFe$_2$Se$_3$ with two objectives. The first is to clarify the microscopic origin of the contrasting electrical resistivity of the two compounds. The second is to examine whether the orbital-ordering pattern proposed from the experimentally observed symmetry lowering [Fig.~\ref{fig:crys}(c)] is supported by first-principles calculations. To this end, we systematically compare candidate magnetic and crystal structures and investigate their electronic structures, orbital ordering, and transport properties. 

\section{Methods}

First-principles calculations were performed within density functional theory (DFT) using the projector augmented-wave method as implemented in VASP~\cite{Kresse1996,Kresse1999}. The exchange--correlation functional was treated within the Perdew--Burke--Ernzerhof generalized gradient approximation~\cite{Perdew1996}. All calculations were performed without spin--orbit coupling using collinear spin-polarized calculations. Correlation effects in the Fe $3d$ orbitals were treated within the rotationally invariant DFT+$U$ scheme with $U=3$~eV and $J=0$~eV.\cite{Dudarev1998} Calculations with $U=0$ were also performed to compare the itinerant and localized regimes. Unless otherwise stated, the electronic structures discussed below were obtained with $U=3$~eV.

The calculations were carried out in two stages for both BaFe$_2$S$_3$ and BaFe$_2$Se$_3$. First, structural optimizations were performed using the experimentally reported $Cmcm$, $Pnma$, and $Pmn2_1$ structures of BaFe$_2$Se$_3$~\cite{Nambu2012,Gao2017,Zheng2020} as common initial structures. The same collinear N\'eel antiferromagnetic configuration was imposed in all three calculations, allowing the intrinsic structural stability and orbital states of the two compounds to be compared independently of differences in magnetic order.

In the second stage, the $Pmn2_1$ structure was adopted as the common starting structure for the N\'eel, Stripe, Block-A, and Block-B magnetic configurations in both compounds. This choice is also motivated by the symmetry description of the block-type antiferromagnetic order. In the $Pnma$ description, the Block AFM structure involves basis functions from separate irreducible representations~\cite{Nambu2012}, whereas it can be represented by a single irreducible representation in $Pmn2_1$~\cite{Aoyama2019}. Appropriate supercells were constructed to accommodate the periodicity of each magnetic order. The crystal symmetry of the relaxed structures was analyzed using the spglib library~\cite{Togo2024} with a symmetry tolerance of \texttt{symprec}=0.03.

The electronic band structures, orbital configurations, and transport properties were analyzed in detail for BaFe$_2$Se$_3$, which exhibits the experimentally observed Block-B magnetic ground state. The Fermi surfaces and electrical-conductivity tensors divided by the relaxation time, $\sigma_{\alpha\beta}/\tau$, were calculated from the interpolated band structures using BoltzTraP2 within semiclassical Boltzmann transport theory and the constant-relaxation-time approximation at $T=300$~K.\cite{Madsen2018} The constant-energy surfaces and momentum-dependent squared velocity component, $v_x^2$, were visualized using FermiSurfer.\cite{Kawamura2019}

$\Gamma$-centered $k$-point meshes were selected according to the size of each magnetic cell. Meshes of $4\times2\times2$ for the N\'eel and Stripe cells and $2\times2\times2$ for the Block-A and Block-B cells were used during structural optimization. The final self-consistent calculations employed denser meshes of $8\times4\times4$, $8\times3\times3$, and $4\times4\times4$ for the N\'eel, Stripe, and Block cells, respectively.

\section{Crystal structure}

To facilitate a direct comparison of the crystal symmetry and the associated orbital degrees of freedom among the $Cmcm$, $Pnma$, and $Pmn2_1$ structures of BaFe$_2$$X_3$, all structures were transformed into a common crystallographic setting based on the orthorhombic $Pmn2_1$ cell. In this setting, the ladder-leg and rung directions are chosen as the $a$ and $c$ axes, respectively. In this setting, the conventional $Cmcm$ and $Pnma$ structures are represented by the equivalent settings $Amam$ and $Pmnb$, respectively.

\begin{figure}
\centering
\includegraphics[width=8cm,pagebox=cropbox,clip]{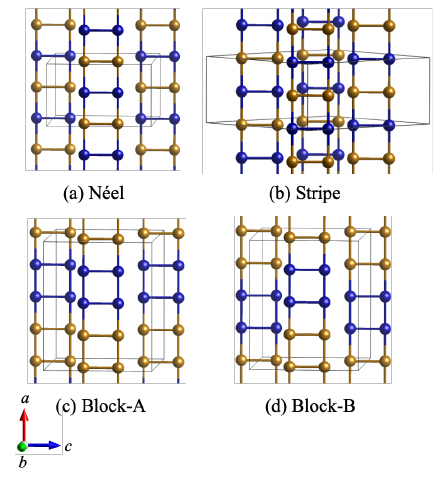}
\caption{
Magnetic configurations considered in this study:
(a) N\'eel, (b) Stripe, (c) Block-A, and (d) Block-B orders.
For clarity, only the Fe sublattice is shown, while Ba and chalcogen (Se/S) atoms are omitted.
Orange and blue spheres represent Fe sites with opposite spin orientations (denoted as Fe and Co in the structural files for technical convenience).
The crystallographic axes are indicated in the lower left corner.
}
\label{fig:ladder}
\end{figure}

To investigate the influence of magnetic order on the structural stability, four representative antiferromagnetic configurations were considered, as illustrated in Fig.~\ref{fig:ladder}. In the N\'eel state [Fig.~\ref{fig:ladder}(a)], nearest-neighbor Fe moments are antiferromagnetically aligned along both the leg ($a$) and rung ($c$) directions of the ladder, and the magnetic structure can be described within the primitive unit cell. In the stripe state [Fig.~\ref{fig:ladder}(b)], Fe moments are ferromagnetically aligned along the rung direction and antiferromagnetically aligned along the leg direction. This state corresponds to the experimentally observed stripe order in BaFe$_2$S$_3$ with propagation vector $\mathbf{q}=(1/2,1/2,0)$,\cite{Takahashi2015,Yu2020} and was modeled using a $\sqrt{2}\times\sqrt{2}\times1$ supercell. 
The Block-A and Block-B states [Figs.~\ref{fig:ladder}(c) and \ref{fig:ladder}(d)] consist of ferromagnetically aligned four-spin blocks ($\uparrow\uparrow\downarrow\downarrow$) along the ladder direction and were described using a $1\times1\times2$ supercell. 
The Block-B state corresponds to the experimentally observed Block order in BaFe$_2$Se$_3$~\cite{Caron2011,Nambu2012}. 
Although the two block states share the same magnetic periodicity along each ladder, they differ in the relative phase of the spin blocks between adjacent ladders, resulting in distinct magnetic symmetries. Structural optimizations including both atomic positions and lattice parameters were subsequently performed for each magnetic configuration.

\section{DFT Results}

\subsection{Structural Stability}

To clarify the intrinsic structural stability of BaFe$_2$S$_3$ and BaFe$_2$Se$_3$, we first compare their crystal structures under a common N\'eel antiferromagnetic configuration. Structural optimizations were performed for both compounds using the experimentally reported $Cmcm$, $Pnma$, and $Pmn2_1$ structures of BaFe$_2$Se$_3$ as the initial structures. Although the low-symmetry $Pnma$ and $Pmn2_1$ structures have not been experimentally reported for BaFe$_2$S$_3$, employing the same set of initial structures enables a direct comparison of their intrinsic structural stability under identical computational conditions. 

For BaFe$_2$S$_3$, the optimized $Cmcm$, $Pnma$, and $Pmn2_1$ structures have relative energies of 134.1, 12.7, and 0~meV/f.u., respectively. Thus, the two low-symmetry structures are energetically much closer to each other than to the high-symmetry $Cmcm$ structure. For BaFe$_2$Se$_3$, the corresponding relative energies are 61.6, 0.94, and 0~meV/f.u., showing that the optimized structures starting from the $Pnma$ and $Pmn2_1$ models become nearly degenerate after structural relaxation. 

These results indicate that the alternating leg-bond disproportionation introduced in the experimental $Pmn2_1$ structure is largely removed during structural relaxation under the N\'eel antiferromagnetic configuration. Consequently, the relaxed $Pmn2_1$ structure becomes very similar to the $Pnma$ structure, particularly in BaFe$_2$Se$_3$. Thus, lowering the crystal symmetry alone is insufficient to stabilize the orbital-ordering pattern proposed for the experimental $Pmn2_1$ structure. In the following subsection, we therefore investigate whether more complex antiferromagnetic configurations, including the Stripe and Block states, can stabilize distinct crystal structures and orbital-ordering patterns.

\subsection{Effects of Magnetic Order on the Structural and Electronic Properties}

\begin{table}[ht]
\caption{
Relative energies $\Delta E$ (meV/f.u.), band gaps $E_{\mathrm{gap}}$ (eV), and resulting space groups (spg) obtained after full structural optimization for different magnetic configurations of BaFe$_2X_3$ ($X=\mathrm{S}$ and Se) at different values of $U$. 
}
\label{tab:energy_gap_spg}
\centering
\begin{tabular}{cc|cccc}
\multicolumn{6}{c}{\textbf{(a)} $U=0$ eV} \\
\hline \hline
$X$ &  & N\'eel & Stripe & Block-A & Block-B \\
\hline

\multirow{3}{*}{S}
& $\Delta E$        & 13.3 & 0 & 162.6 & 140.0 \\
& $E_{\rm gap}$     & 0.16 & 0.13 & 0.33 & 0.42 \\
& spg               & $Pnma$ & $Pnma$ & $Pmc2_1$ & $Pc$ \\
\hline

\multirow{3}{*}{Se}
& $\Delta E$        & 60.3 & 71.7 & 14.5 & 0 \\
& $E_{\rm gap}$     & 0.09 & 0.03 & 0.37 & 0.47 \\
& spg               & $Pnma$ & $Cmcm$ & $Pm$ & $P2_1$ \\
\hline\hline
\\

\multicolumn{6}{c}{\textbf{(b)} $U=3$ eV} \\
\hline \hline
$X$ &  & N\'eel & Stripe & Block-A & Block-B \\
\hline

\multirow{3}{*}{S}
& $\Delta E$        & 59.9 & 12.2 & 19.8 & 0 \\
& $E_{\rm gap}$     & 1.13 & 1.26 & 1.19 & 1.25 \\
& spg               & $Pnma$ & $P2_1/c$ & $Pmc2_1$ & $Pc$ \\
\hline

\multirow{3}{*}{Se}
& $\Delta E$        & 38.9 & 77.5 & 33.2 & 0 \\
& $E_{\rm gap}$     & 1.22 & 1.07 & 1.07 & 1.23 \\
& spg               & $Pnma$ & $P2_1/c$ & $Pmc2_1$ & $Pc$ \\
\hline\hline
\end{tabular}
\end{table}

Since the N\'eel configuration does not stabilize the proposed orbital-ordering pattern, we next examine the effects of different antiferromagnetic configurations on the structural stability, orbital ordering, and electronic properties of BaFe$_2$S$_3$ and BaFe$_2$Se$_3$. The calculated total energies, band gaps, and optimized space groups are summarized in Table~\ref{tab:energy_gap_spg}.

For BaFe$_2$S$_3$, the GGA ($U=0$) calculations reproduce the experimentally observed Stripe antiferromagnetic ground state, although the optimized structure has $Pnma$ symmetry rather than the experimentally reported $Cmcm$ structure.\cite{Zheng2018} The calculated band gaps remain smaller than 0.5~eV, and no pronounced symmetry lowering is obtained. In contrast, introducing $U=3$~eV stabilizes the Block-B state with a lower-symmetry structure, which is inconsistent with the experimental magnetic ground state.

For BaFe$_2$Se$_3$, the situation is different. Within the GGA+$U$ ($U=3$~eV) calculations, the Block-B configuration becomes the lowest-energy magnetic state and is accompanied by the low-symmetry $Pc$ structure, consistent with the experimentally observed Block antiferromagnetic order.\cite{Caron2011,Nambu2012} By contrast, the GGA calculations favor relatively high-symmetry structures with small band gaps and do not reproduce the experimentally proposed structural distortion.

These results suggest that BaFe$_2$Se$_3$ may have a stronger tendency toward electron localization than BaFe$_2$S$_3$. Such a tendency is qualitatively consistent with the larger lattice constant of BaFe$_2$Se$_3$, which is expected to reduce the effective hopping between Fe ions. Correspondingly, the experimentally observed magnetic and structural properties of BaFe$_2$S$_3$ are reproduced more successfully within the GGA calculations, whereas those of BaFe$_2$Se$_3$ are better reproduced by GGA+$U$.

\begin{figure}[th]
\centering
\includegraphics[width=\columnwidth,pagebox=cropbox,clip]{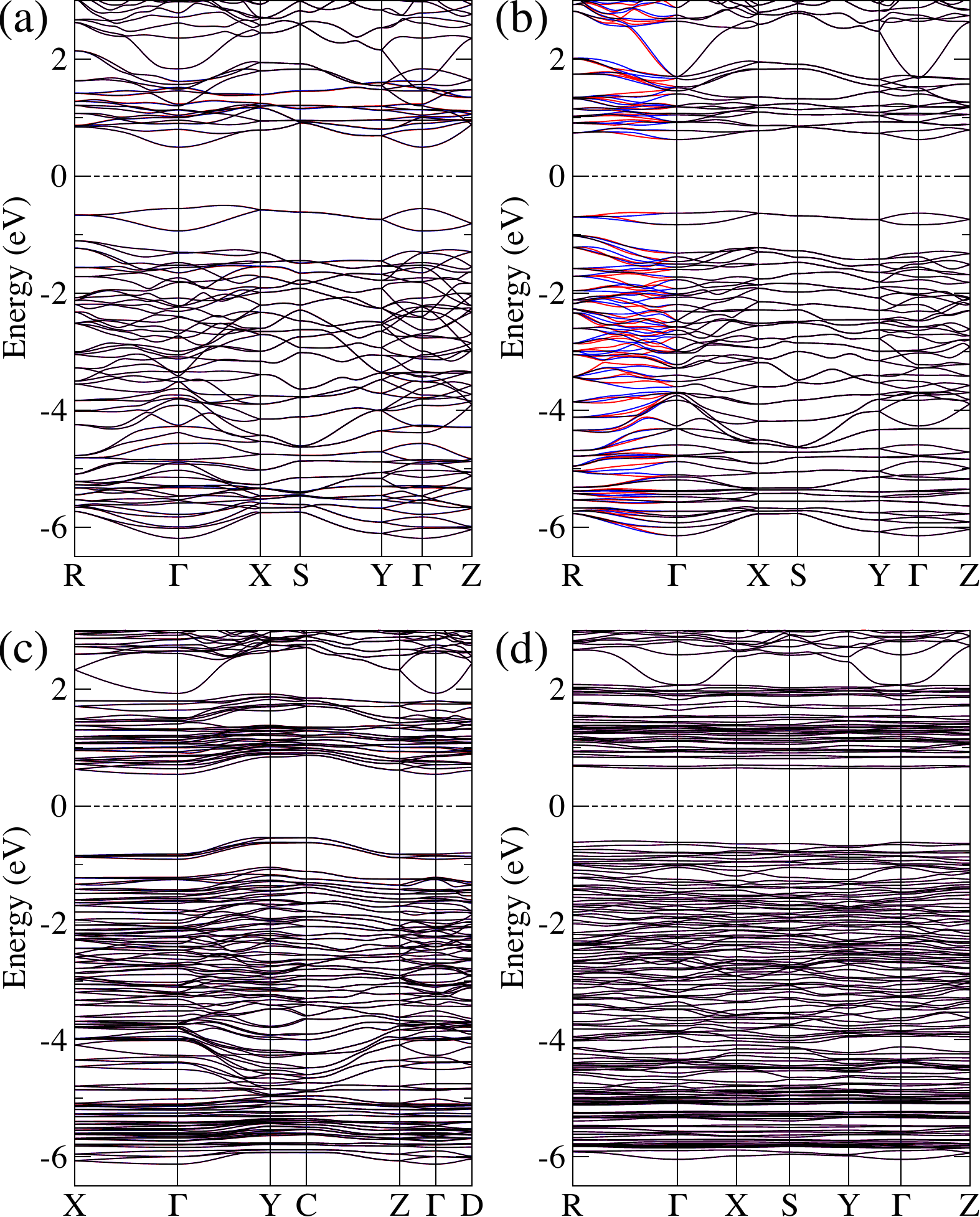}
\caption{
Electronic band structures of the optimized
(a) $Cmcm$ N\'eel,
(b) $Pnma$ N\'eel, 
(c) $P2_1/c$ Stripe,
and
(d) $Pc$ Block-B states obtained under the respective fixed magnetic
orders in BaFe$_2$Se$_3$. 
Spin--orbit coupling was not included.
The up- and down-spin bands are distinguished by red and blue curves. 
}
\label{fig:band}
\end{figure}

Having established the different magnetic ground states of BaFe$_2$S$_3$ and BaFe$_2$Se$_3$, we now focus on BaFe$_2$Se$_3$ as a representative system to examine how the magnetic configuration modifies the electronic structure. Figure~\ref{fig:band} compares the spin-resolved electronic band structures of the optimized states. 

In the high-symmetry $Cmcm$ N\'eel state [Fig.~\ref{fig:band}(a)], the up- and down-spin bands remain degenerate throughout the Brillouin zone. This complete spin degeneracy results from the combination of the crystallographic symmetry and the N\'eel magnetic order. When the crystal symmetry is lowered from $Cmcm$ to $Pnma$ while retaining the same N\'eel magnetic configuration [Fig.~\ref{fig:band}(b)], a pronounced momentum-dependent splitting between the up- and down-spin bands appears, particularly along the $R$--$\Gamma$ direction. Such a spin splitting in the absence of spin--orbit coupling and a net ferromagnetic moment is characteristic of an altermagnetic band structure. 

In contrast, the $P2_1/c$ Stripe and $Pc$ Block-B states [Figs.~\ref{fig:band}(c) and (d)] exhibit only weak spin splitting despite their lower crystal symmetry. Although symmetry does not generally prohibit spin splitting in these magnetic structures, the calculated results indicate that the primary effect of the Stripe and Block magnetic orders is to modify the band dispersions and insulating gaps rather than to produce the pronounced momentum-dependent spin splitting found in the $Pnma$ N\'eel state.

\begin{figure}
\centering
\includegraphics[width=0.8\columnwidth,pagebox=cropbox,clip]{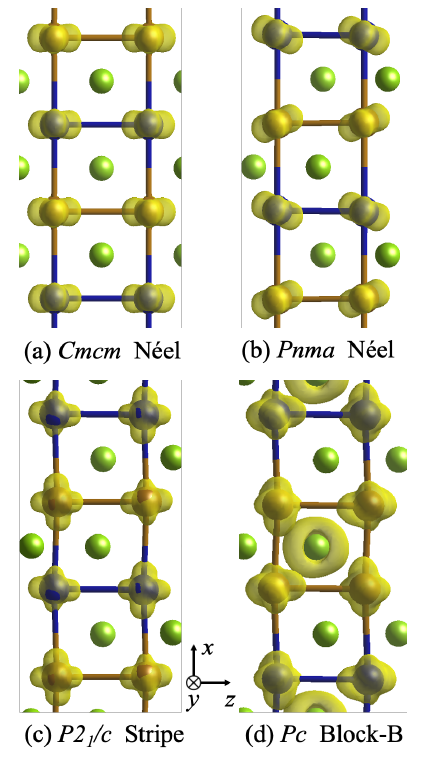}
\caption{
Partial charge densities integrated over the energy range from
0.5~eV below the valence-band maximum to the valence-band maximum in BaFe$_2$Se$_3$. 
The crystal symmetry is followed by the magnetic order:
(a) $Cmcm$, N\'eel,
(b) $Pnma$, N\'eel,
(c) $P2_1/c$, Stripe,
and (d) $Pc$, Block-B.
}
\label{fig:orbital}
\end{figure}

Figure~\ref{fig:orbital} presents the partial charge densities integrated over the energy range from 0.5~eV below the valence-band maximum to the valence-band maximum, providing a direct visualization of the occupied orbital states. Since Fe$^{2+}$ ($d^6$) ions in tetrahedral coordination have one electron occupying the doubly degenerate $e$ orbitals, the orbital degree of freedom plays an essential role in the electronic structure. The partial charge densities are shown for the optimized structures obtained by relaxing each crystal symmetry under the corresponding magnetic order.

In the $Cmcm$ N\'eel state [Fig.~\ref{fig:orbital}(a)], all Fe sites exhibit nearly identical orbital characters, with the occupied ${y^2-z^2}$-like orbitals lying within planes perpendicular to the ladder leg direction. This ferroic arrangement reflects the high crystallographic symmetry of the $Cmcm$ structure.

In the $Pnma$ N\'eel state [Fig.~\ref{fig:orbital}(b)], the canting of the ladders induces a modulation of the orbital orientation. The ${y^2-z^2}$-like orbitals on the upper two Fe sites are tilted in one direction, whereas those on the lower two Fe sites are tilted in the opposite direction, resulting in an alternating orbital pattern within the ladder. This behavior demonstrates that the orbital degrees of freedom are strongly coupled to the lattice distortion.

In the $P2_1/c$ Stripe state [Fig.~\ref{fig:orbital}(c)], the reduction of crystal symmetry further lifts the equivalence of the Fe pairs along the ladder leg. As a result, the occupied orbitals develop an antiferroic $3x^2-r^2/3z^2-r^2$ orbital arrangement, consistent with the orbital-ordering model proposed for the experimentally suggested $Pmn2_1$ structure [Fig.~\ref{fig:crys}(c)]. This orbital pattern is analogous to the cooperative Jahn--Teller orbital ordering well known in perovskite manganites.

In the $Pc$ Block-B state [Fig.~\ref{fig:orbital}(d)], the occupied Fe orbitals also exhibit an antiferroic $3x^2-r^2/3z^2-r^2$ orbital arrangement. In addition, the four Fe sites within each block have the same spin orientation, allowing their occupied orbitals to strongly hybridize with the central Se $p$ orbitals and form a cluster-like molecular orbital extending over the entire Fe$_4$Se unit. The accompanying symmetry lowering further enhances the orbital polarization through a cooperative Jahn--Teller-like distortion, thereby increasing the band gap and stabilizing the $Pc$ state. A detailed comparison of the orbital states in the Block-type AFM configurations is provided in Appendix~\ref{app:blockb_orbitals}.

The orbital-ordering patterns obtained for the four antiferromagnetic configurations indicate that the cooperative Jahn--Teller-type orbital ordering schematically illustrated in Fig.~\ref{fig:crys}(c) is not stabilized simply by adopting the proposed $Pmn2_1$ crystal structure. Instead, it emerges only in the $P2_1/c$ Stripe and $Pc$ Block-B states, where the ladders develop a staggered distortion perpendicular to the ladder direction. This additional structural degree of freedom couples to the cooperative Jahn--Teller distortion and stabilizes the antiferroic orbital ordering. Furthermore, the $Pc$ Block-B state exhibits additional Fe$_4$Se cluster-like molecular orbitals, distinguishing it from the Stripe state. 

The orbital ordering provides a possible explanation for the contrasting transport properties of the experimentally observed phases of BaFe$_2$S$_3$ and BaFe$_2$Se$_3$. In the $Cmcm$ Stripe state realized experimentally in BaFe$_2$S$_3$, the occupied states are expected to retain predominantly ${y^2-z^2}$ character. Consequently, substantial Fe--S--Fe hopping channels remain active. In contrast, the $Pc$ Block-B state of BaFe$_2$Se$_3$ exhibits a qualitatively different electronic structure. The occupied states form Fe$_4$--Se cluster orbitals, within which the Fe $3d$ and Se $p$ states are strongly hybridized. Furthermore, the antiferroic orbital ordering suppresses the inter-cluster hopping amplitudes. As a result, the electronic states become spatially confined within the Fe$_4$Se clusters, leading to a larger resistivity in BaFe$_2$Se$_3$ than in BaFe$_2$S$_3$. The present results therefore suggest that the contrasting transport behaviors of the two compounds originate from fundamentally different orbital and bonding characters established through the interplay of magnetic order, orbital ordering, and Fe$_4$Se cluster formation.

\subsection{Electrical Conductivity from Boltzmann Transport}

\begin{figure}[t]
\centering
\includegraphics[
  width=\columnwidth,
  pagebox=cropbox,
  clip
]{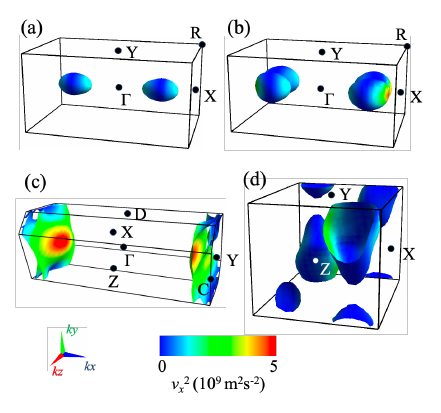}
\caption{
Constant-energy surfaces at $E = E_{\mathrm{v}}-0.02$~eV for the
(a) $Cmcm$ N\'eel,
(b) $Pnma$ N\'eel,
(c) $P2_1/c$ Stripe, and
(d) $Pc$ Block-B states in BaFe$_2$Se$_3$. 
The color represents the squared group-velocity component
$v_x^2$, where the $x$ direction corresponds to the ladder-leg
direction. 
The high-symmetry points are labeled using the orthorhombic convention
for the N\'eel and Block-B states, and the monoclinic
convention for the Stripe state. 
}
\label{fig:fermisurfaces}
\end{figure}

To investigate how the different magnetic configurations affect the transport properties, we calculated the semiclassical electrical conductivity using the BoltzTraP2 code at $T=300$ K. 
Since the absolute conductivity depends on the unknown carrier relaxation time, we focus on the anisotropy of the conductivity tensor. The conductivity was evaluated at representative energies near the band edges. These energies correspond to moderate hole and electron doping levels while avoiding the numerical instability exactly at the band edges. 
Within the constant-relaxation-time approximation, the diagonal conductivity is determined by the velocity-squared component weighted over the states near the chemical potential,
\begin{equation}
\frac{\sigma_{\alpha\alpha}}{\tau}
=
e^2 \sum_n \int_{\rm BZ}\frac{d\mathbf{k}}{(2\pi)^3}
v_{n\alpha}^2(\mathbf{k})
\left(-\frac{\partial f}{\partial \varepsilon}\right)
_{\varepsilon=\varepsilon_{n\mathbf{k}}},
\end{equation}
where $v_{n\alpha}(\mathbf{k})=\hbar^{-1}\partial\varepsilon_{n\mathbf{k}}/\partial k_\alpha$. 
Accordingly, electronic states with large $v_x^2$ enhance the ladder-leg conductivity $\sigma_{xx}/\tau$.

\begin{table}[t]
\caption{Electrical conductivity in BaFe$_2$Se$_3$ divided by the relaxation time,
$\sigma/\tau$, at $T=300$ K.
Here, $E_{\mathrm{v}}-\Delta E$ and $E_{\mathrm{c}}+\Delta E$
represent energies $\Delta E$ (eV) below the valence-band maximum and
above the conduction-band minimum, respectively.
The unit of $\sigma/\tau$ is
$10^{19}\,\Omega^{-1}\,\mathrm{m}^{-1}\,\mathrm{s}^{-1}$.
The $x$, $y$, and $z$ directions correspond to the ladder leg direction,
the direction normal to the ladder plane, and the ladder rung direction,
respectively.}
\label{tab:conductivity}
\centering
\begin{tabular}{llccc}
\hline\hline
State & Energy & $\sigma_{xx}/\tau$ & $\sigma_{yy}/\tau$ & $\sigma_{zz}/\tau$ \\
\hline
$Cmcm$ N\'eel
& $E_{\mathrm{c}}+0.2$ & 4.51 & 3.66 & 3.26 \\
& $E_{\mathrm{c}}+0.1$ & 4.83 & 3.67 & 3.05 \\
& $E_{\mathrm{v}}-0.1$ & 4.11 & 2.39 & 1.71 \\
& $E_{\mathrm{v}}-0.2$ & 6.42 & 1.55 & 2.34 \\
\hline
$Pnma$ N\'eel
& $E_{\mathrm{c}}+0.2$ & 3.28 & 2.66 & 1.55 \\
& $E_{\mathrm{c}}+0.1$ & 3.04 & 2.47 & 1.50 \\
& $E_{\mathrm{v}}-0.1$ & 2.16 & 1.28 & 1.09 \\
& $E_{\mathrm{v}}-0.2$ & 5.53 & 1.64 & 1.19 \\
\hline
$P2_1/c$ Stripe
& $E_{\mathrm{c}}+0.2$ & 0.87 & 0.79 & 1.08 \\
& $E_{\mathrm{c}}+0.1$ & 0.75 & 0.89 & 0.91 \\
& $E_{\mathrm{v}}-0.1$ & 5.61 & 0.25 & 0.30 \\
& $E_{\mathrm{v}}-0.2$ & 7.06 & 0.29 & 0.29 \\
\hline
$Pc$ Block-B
& $E_{\mathrm{c}}+0.2$ & 0.62 & 0.38 & 1.51 \\
& $E_{\mathrm{c}}+0.1$ & 0.85 & 0.41 & 2.13 \\
& $E_{\mathrm{v}}-0.1$ & 1.28 & 0.45 & 0.96 \\
& $E_{\mathrm{v}}-0.2$ & 1.70 & 0.32 & 0.49 \\
\hline\hline
\end{tabular}
\end{table}

Table~\ref{tab:conductivity} summarizes the electrical conductivity divided by the relaxation time for the four magnetic configurations.
The $Cmcm$ and $Pnma$ N\'eel states and the $P2_1/c$ Stripe state exhibit
enhanced ladder-leg transport on the hole-doped side.  Cooperative alignment of the occupied $e^1$ orbitals within the ladder plane forms an effective hopping channel along the ladder leg, yielding quasi-one-dimensional constant-energy surfaces with large $v_x^2$, as shown in Fig.~\ref{fig:fermisurfaces}.  In the $Pnma$ N\'eel state, the hole pocket is split into two pockets along the $k_z$ direction as a consequence of the altermagnetic band splitting discussed in the previous subsection [Fig.~\ref{fig:fermisurfaces}(b)]. Despite this splitting, both pockets retain large $v_x^2$, and the ladder-leg conductivity therefore remains enhanced. 
The hole-side conductivity is most anisotropic in the Stripe state, for which $\sigma_{xx}/\tau$ exceeds the transverse components by more
than one order of magnitude.

The $Pc$ Block-B state shows a different transport character. Its hole-side conductivity is largest along the ladder leg, whereas its electron-side conductivity is dominated by the $z$ component. Nevertheless, its ladder-leg hole conductivity is substantially smaller than that of the Stripe state. As shown in Table~\ref{tab:conductivity}, at $E_{\mathrm{v}}-0.10$ and $-0.20$~eV, $\sigma_{xx}/\tau$ is $5.61$ and $7.06\times10^{19}\,\Omega^{-1}\,\mathrm{m}^{-1}\,\mathrm{s}^{-1}$, respectively, for the Stripe state, compared with $1.28$ and $1.70\times10^{19}\,\Omega^{-1}\,\mathrm{m}^{-1}\,\mathrm{s}^{-1}$ for the Block-B state. This suppression is consistent with the confinement of electronic states within magnetic Fe$_4$ units by the Block-type antiferromagnetic order, which suppresses coherent interblock hopping. 
Although the carrier type in the ambient-pressure Mott-insulating state has not been established experimentally, Hall measurements in the pressure-induced metallic phase indicate hole-dominant transport near the superconducting region~\cite{Aoyama2026}. We therefore discuss the hole-doped results as a relevant reference for the transport anisotropy.

Although the present calculations do not directly describe the orbital-state changes across $T^\ast$, because this transition occurs above $T_{\rm N}$ where long-range antiferromagnetic order is absent, they demonstrate that distinct antiferromagnetic configurations stabilize distinct orbital-ordering patterns, leading to markedly different transport properties. The comparison between the $Cmcm$ and $Pnma$ N\'eel states further indicates that even a change between two ferroic orbital-ordering patterns modifies the ladder-leg conductivity, highlighting the important role of the orbital state in transport. Assuming the experimentally observed Stripe and Block-B magnetic ground states for BaFe$_2$S$_3$ and BaFe$_2$Se$_3$, respectively, the substantially lower hole conductivity of the Block-B state provides a qualitative explanation for the approximately two-orders-of-magnitude higher resistivity of BaFe$_2$Se$_3$. Overall, the present calculations provide a microscopic picture of how magnetic order, orbital ordering, and transport properties are coupled through the systematic comparison of candidate magnetic and crystal structures while consistently reproducing the Mott-insulating state. 
Furthermore, the present calculations suggest that the Block-B magnetic ground state is accompanied by a low-symmetry $Pc$ structure, and experimental verification of this proposed structural scenario will provide further insight into the coupling between magnetic order and lattice distortion.

\section{Summary}

We investigated the interplay among magnetic order, structural stability, orbital ordering, and transport anisotropy in BaFe$_2$S$_3$ and BaFe$_2$Se$_3$ using first-principles calculations. The optimized structures depend strongly on the magnetic configuration. Within bare GGA, the Stripe state is energetically favored for BaFe$_2$S$_3$, whereas GGA+$U$ stabilizes the Block-B state together with a lower-symmetry crystal structure. These results reproduce the experimentally observed tendencies toward Stripe antiferromagnetism in BaFe$_2$S$_3$ and Block-B antiferromagnetism in BaFe$_2$Se$_3$, although the optimized crystal symmetries differ from the currently proposed structural models. In particular, the calculations predict that the Block-B magnetic ground state in BaFe$_2$Se$_3$ is accompanied by a lower-symmetry $Pc$ structure, providing a new structural signature to be examined experimentally. 
The calculations further clarify the microscopic origin of the orbital states. The antiferroic orbital-ordering pattern proposed for the experimental $Pmn2_1$ structure is not stabilized under the simple N\'eel antiferromagnetic configuration. Instead, the Block-B state exhibits a different antiferroic orbital ordering accompanied by strong Fe $3d$--Se $p$ hybridization, resulting in Fe$_4$Se cluster-like molecular orbitals. The calculated transport anisotropy shows that the Block-B state has substantially lower hole conductivity along the ladder-leg direction than the Stripe state, providing a qualitative explanation for the much higher resistivity of BaFe$_2$Se$_3$. These results demonstrate how magnetic order governs the structural stability, orbital ordering, and transport properties of iron-ladder compounds through the coupled orbital--lattice degrees of freedom.

\begin{acknowledgments}
The present work was financially supported by JSPS KAKENHI Grant Numbers 
JP18H01159, 
JP20K14396, 
JP25K00955 
and 
supported by JST-CREST (Grant No. JPMJCR22O2) and by
Institute for Open and Transdisciplinary Research Initiatives (OTRI), the University of Osaka.
The computation in this work has been done using the facilities of the Supercomputer Center, the Institute for Solid State Physics, the University of Tokyo. The crystallographic figure was generated using the VESTA program~\cite{Momma2011}. 
\end{acknowledgments}

\appendix

\section{Crystallographic Symmetry Analysis}
\label{app:symmetry}

We summarize the crystallographic symmetry operations used to compare the $Cmcm$, $Pnma$, and $Pmn2_1$ structures of BaFe$_2$Se$_3$. In this Appendix, we consider only the crystallographic space-group symmetry of the crystal structure, without including spin degrees of freedom or spin-orbit coupling. All structures are expressed in a common orthorhombic setting based on the $Pmn2_1$ cell, in which the ladder-leg and rung directions are chosen as the $a$ and $c$ axes, respectively. In this setting, the conventional $Cmcm$ and $Pnma$ structures are represented by the equivalent $Amam$ and $Pmnb$ settings, respectively. This common setting makes explicit how the Fe crystallographic orbit evolves as the symmetry is lowered.

In the $Amam$ setting (No.~63), the symmetry operations are
\begin{align}
G_{Amam}=\{&
\{E|000\},
\{C_{2z}|000\},
\{C_{2y}|\tfrac12 00\},
\{C_{2x}|\tfrac12 00\}, \nonumber\\
&
\{I|000\},
\{m_z|000\},
\{m_y|\tfrac12 00\},
\{m_x|\tfrac12 00\}, \nonumber\\
&
\{E|0\tfrac12\tfrac12\},
\{C_{2z}|0\tfrac12\tfrac12\},
\{C_{2y}|\tfrac12\tfrac12\tfrac12\},
\{C_{2x}|\tfrac12\tfrac12\tfrac12\}, \nonumber\\
&
\{I|0\tfrac12\tfrac12\},
\{m_z|0\tfrac12\tfrac12\},
\{m_y|\tfrac12\tfrac12\tfrac12\},
\{m_x|\tfrac12\tfrac12\tfrac12\}
\}.
\end{align}
Because of these operations and the A-centering translation, all Fe atoms belong to a single crystallographic orbit, Fe1 $8e$.

In the $Pmnb$ setting (No.~62), the symmetry operations are
\begin{align}
G_{Pmnb}=\{&
\{E|000\},
\{C_{2y}|\tfrac12\tfrac12\tfrac12\},
\{C_{2x}|\tfrac12 00\},
\{C_{2z}|0\tfrac12\tfrac12\}, \nonumber\\
&
\{I|000\},
\{m_z|\tfrac12\tfrac12\tfrac12\},
\{m_y|\tfrac12 00\},
\{m_x|0\tfrac12\tfrac12\}
\}.
\end{align}
Although the centering translation is absent and the ladder geometry is lower in symmetry than in $Amam$, inversion symmetry remains. All Fe atoms therefore still belong to a single crystallographic orbit, Fe1 $8d$.

In the $Pmn2_1$ structure (No.~31), the symmetry operations reduce to
\begin{align}
G_{Pmn2_1}=\{&
\{E|000\},
\{C_{2z}|\tfrac12 0 \tfrac12\},
\{m_x|000\},
\{m_y|\tfrac12 0 \tfrac12\}
\}.
\end{align}
Inversion symmetry is absent in this structure. As a result, the single Fe orbit present in $Amam$ and $Pmnb$ splits into two crystallographically distinct Fe sublattices, Fe1 $4b$ and Fe2 $4b$.

Thus, the essential symmetry distinction is not the reduction from $Cmcm$ ($Amam$) to $Pnma$ ($Pmnb$), where the Fe sublattice remains a single crystallographic orbit, but the further reduction to $Pmn2_1$, where inversion symmetry is lost and the Fe sublattice splits into two inequivalent crystallographic sites. This Fe-site splitting provides the crystallographic basis for site-dependent orbital polarization.

\begin{figure}[t]
\centering
\includegraphics[width=0.6\columnwidth]{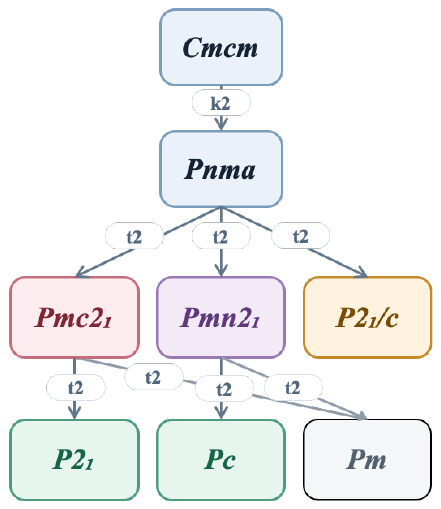}
\caption{
Group--subgroup relations among the space groups relevant to BaFe$_2$Se$_3$.
The labels $t2$ and $k2$ denote translationengleiche and klassengleiche subgroup relations of index 2, respectively. The diagram summarizes symmetry-allowed relations among the structures considered in this work and does not represent a unique structural
transition pathway.  
}
\label{fig:space_group_relations}
\end{figure}

Figure~\ref{fig:space_group_relations} summarizes the group--subgroup relations among the space groups relevant to the present study.
The relation from $Cmcm$ to $Pnma$ is a k2 relation: both structures have the orthorhombic point group $mmm$, whereas the C-centering translation of $Cmcm$ is lost in $Pnma$.
In contrast, the t2 relations from $Pnma$ to $Pmn2_1$,
$Pmc2_1$, and $P2_1/c$ preserve the translation lattice while reducing
the point-group symmetry.
The $P2_1$ and $Pc$ structures obtained for the Block-B state at $U=0$ and $3$~eV, respectively, are further low-symmetry subgroups.
The reported $Pm$ structure is also included as a proposed low-symmetry experimental candidate.\cite{Zheng2020}

\section{Ladder Structure Analysis}
\label{app:ladder_structure}

\begin{table}[ht]
\caption{
Representative short and long leg bond lengths
($d_{\rm short}$ and $d_{\rm long}$),
rung bond length ($r_{\rm rung}$),
and leg tilt angle ($\theta$)
for the optimized magnetic states.
The Block-A state is characterized by a strong intra-/inter-block
bond disproportionation while maintaining an almost straight ladder
($\theta \approx 0^\circ$),
whereas the Block-B state exhibits a pronounced zigzag distortion
with finite leg tilt angles.
}
\label{tab:ladder}
\begin{ruledtabular}
\begin{tabular}{lcccc}
\hline
State &
$d_{\rm short}$ (\AA) &
$d_{\rm long}$ (\AA) &
$r_{\rm rung}$ (\AA) &
$\theta$ (deg) \\
\hline
Block-A & 2.57--2.58 & 2.89--2.91 & 2.70 & 0.0 \\
Block-B & 2.62--2.64 & 2.86--2.88 & 2.87--2.88 & 2.3--2.7 \\
N\'eel & 2.80 & 2.80 & 2.98 & 2.80--2.85 \\
Stripe & 2.82--2.83 & 2.86--2.87 & 2.90 & 2.8--3.0 \\
\hline
\end{tabular}
\end{ruledtabular}
\end{table}

The structural characteristics of the optimized magnetic states in BaFe$_2$Se$_3$ 
are summarized in Table~\ref{tab:ladder}.
The Block-A state exhibits the largest leg-bond disproportionation,
with short and long leg bonds of approximately 2.57 and 2.90~\AA,
respectively.
Despite this pronounced intra-/inter-block bond modulation,
the leg bonds remain nearly parallel to the ladder direction,
yielding a negligible tilt angle ($\theta \approx 0^\circ$).
Thus, the essential structural feature of Block-A is the formation
of Fe$_4$ blocks on an almost straight ladder.

In contrast, the Block-B state is characterized by a substantial
zigzag distortion of the ladder.
Although the difference between short and long leg bonds remains
significant (2.62--2.64 versus 2.86--2.88~\AA),
the finite leg tilt angles ($\theta = 2.3$--$2.7^\circ$)
indicate that the ladder itself becomes buckled.
Therefore, the primary distinction between Block-A and Block-B
lies not only in the bond disproportionation but also in the
presence or absence of the zigzag ladder distortion.

The N\'eel state exhibits nearly uniform leg bonds
($d_{\rm short}=d_{\rm long}\approx2.80$~\AA)
together with the longest rung bonds
($r_{\rm rung}\approx2.98$~\AA),
suggesting a relatively undistorted ladder geometry.
The Stripe state shows a moderate leg-bond disproportionation
($d_{\rm short}\approx2.82$~\AA,
$d_{\rm long}\approx2.87$~\AA)
and finite tilt angles comparable to those of the N\'eel state.

The relationship between these structural distortions and the orbital ordering is discussed in Appendix C.

\section{Comparison of Orbital Ordering Patterns with Block-type AFM Configuration}
\label{app:blockb_orbitals}

\begin{table}[ht]
\caption{
Relative energies $\Delta E$ (meV/f.u.), band gaps $E_{\mathrm{gap}}$ (eV), and resulting space groups (spg) obtained after full structural optimization of BaFe$_2$Se$_3$ with the Block-A and Block-B magnetic configurations starting from the experimental $Pm$ and $Pmn2_1$ structures.\cite{Zheng2020} The lowest-energy state is highlighted in bold.
}
\label{tab:initial_structure}
\centering
\begin{tabular}{cc|cc}
\hline\hline
Initial  
structure &  & Block-A & Block-B \\
\hline

\multirow{3}{*}{$Pm$}
& $\Delta E$        & 48.2 & 37.8 \\
& $E_{\rm gap}$     & 1.05 & 1.06 \\
& spg               & $Pmc2_1$ & $Pmn2_1$ \\
\hline

\multirow{3}{*}{$Pmn2_1$}
& $\Delta E$        & 33.2 & \textbf{0} \\
& $E_{\rm gap}$     & 1.07 & 1.23 \\
& spg               & $Pmc2_1$ & $Pc$ \\
\hline\hline
\end{tabular}
\end{table}

In the x-ray diffraction analysis by Zheng \textit{et al.},\cite{Zheng2020} both $Pmn2_1$ and $Pm$ structures were proposed for BaFe$_2$Se$_3$. To examine the stability of these structures, we performed structural relaxations for both the Block-A and Block-B magnetic configurations starting from each experimental structure. The results are summarized in Table~\ref{tab:initial_structure}. Starting from the experimental $Pm$ structure, the optimized structures become $Pmc2_1$ for the Block-A state and $Pmn2_1$ for the Block-B state, indicating that the crystal relaxes toward higher-symmetry structures. In contrast, starting from the experimental $Pmn2_1$ structure, the Block-B state lowers its symmetry to $Pc$. This symmetry lowering is accompanied by an increase in the band gap from 1.07 to 1.23~eV, suggesting that the structural distortion is coupled to orbital ordering, as discussed below.

\begin{figure}[ht]
\centering
\includegraphics[width=1.0\columnwidth]{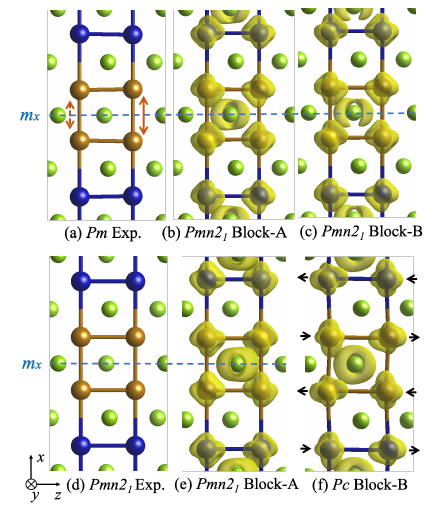}
\caption{
Comparison of the experimental structures and the partial charge densities integrated over the energy window $E_{\rm VBM}-0.5~\mathrm{eV}<E<E_{\rm VBM}$. (a) Experimental $Pm$ structure.\cite{Zheng2020} (b) and (c) Partial charge densities obtained after structural relaxation starting from the experimental $Pm$ structure in (a) with the Block-A and Block-B magnetic configurations, respectively. (d) Experimental $Pmn2_1$ structure.\cite{Zheng2020} (e) and (f) Partial charge densities obtained after structural relaxation starting from the experimental $Pmn2_1$ structure in (d) with the Block-A and Block-B magnetic configurations, respectively. The blue dashed lines indicate the $m_x$ mirror planes. The orange arrows in (a) indicate the alternating long and short Fe--Fe bonds along the ladder legs in the experimental $Pm$ structure. The black arrows in (f) indicate the staggered ladder distortion characteristic of the $Pc$ Block-B structure.
}
\label{fig:orbital_comparison}
\end{figure}

Figure~\ref{fig:orbital_comparison} summarizes the orbital states obtained after structural relaxation with the Block-A and Block-B magnetic configurations starting from the experimentally proposed $Pm$ and $Pmn2_1$ structures.\cite{Zheng2020} The experimentally proposed $Pm$ structure with alternating long and short Fe--Fe bonds along the ladder legs does not reproduce the orbital ordering proposed by Aoyama \textit{et al.} shown in Fig.~\ref{fig:crys}(c). Instead, structural relaxation removes the leg-bond disproportionation for both the Block-A and Block-B states, yielding the higher-symmetry $Pmc2_1$ and $Pmn2_1$ structures, respectively. This indicates that the experimentally proposed $Pm$ distortion is not a local minimum within the present GGA+$U$ calculations. Although the magnetic Fe$_4$Se blocks remain intact, the orbital ordering expected from the experimental $Pm$ model is not reproduced.

To understand the origin of the orbital ordering, it is useful to consider the symmetry operations of the parent $Pmn2_1$ structure,
\begin{align}
G_{\rm Pmn2_1}=\{&
\{E|000\},
\{C_{2z}|\tfrac12 0 \tfrac12\},
\{m_x|000\},
\{m_y|\tfrac12 0 \tfrac12\}
\},
\end{align}
where $m_x$ is the intra-ladder mirror operation, whereas $\{C_{2z}|\tfrac12 0 \tfrac12\}$ and $\{m_y|\tfrac12 0 \tfrac12\}$ relate neighboring ladders. 
The essential difference between the Block-A and Block-B magnetic structures is whether the intra-ladder mirror symmetry $m_x$ is preserved. In the Block-A state, the Block antiferromagnetic patterns on neighboring ladders are in phase. The intra-ladder mirror operation $m_x$ therefore transforms each ladder into an equivalent magnetic configuration, and the optimized crystal structure retains this mirror symmetry, resulting in the $Pm$ space group. In contrast, the Block-B state exhibits a half-period phase shift between neighboring ladders. Under the $m_x$ operation, this phase-shifted magnetic pattern is no longer mapped onto itself, and the mirror symmetry is therefore broken. 
Consequently, the crystal structure is allowed to undergo an additional symmetry lowering to the $Pc$ space group.
This additional structural degree of freedom enables the staggered distortion perpendicular to the ladder direction, which couples to the cooperative Jahn--Teller distortion, enhancing the orbital polarization associated with the Fe \(e\) states and thereby further stabilizing the cooperative orbital ordering. 

\bibliographystyle{apsrev4-2}
\bibliography{reference}

\end{document}